\documentclass[11pt,a4paper]{article}
\pdfoutput=1
\usepackage{jheppub}
\usepackage[hyperref,svgnames]{xcolor}
\usepackage[latin1]{inputenc}
\usepackage{amsmath}
\usepackage{amsfonts}
\usepackage{amssymb}
\usepackage{booktabs}
\usepackage{graphicx}
\usepackage{setspace}
\usepackage{mathrsfs}
\usepackage{tikz}
\usepackage{caption}
\usepackage{subcaption}

\definecolor{hgreen}{rgb}{0,.3,0}
\definecolor{hred}{rgb}{.3,0,0}
\definecolor{hblue}{rgb}{0,0,.3}
\definecolor{LightGray}{gray}{0.95}

\numberwithin{equation}{section}

\title{Multi-Field Q-balls with Real Scalars\\
}

\author{Olivier Lennon}
\emailAdd{olivier.lennon@physics.ox.ac.uk}

\affiliation{Rudolf Peierls Centre for Theoretical Physics, University of Oxford, Clarendon Laboratory, Parks Road, Oxford OX1 3PU, United Kingdom}

\abstract{
Multi-field Q-balls, in which some, but not all, of the constituent fields are real scalars, are studied. Uncharged fields may classically contribute to Q-balls provided that their effect is to not destabilise the resulting object. The thin-wall limit is reviewed based on existing literature. Sufficient conditions for the theories that can plausibly contain thick-wall solutions are derived. Necessary conditions require a detailed numerical analysis, outside the scope of this work. These conditions show that when the additional real scalars are massive, a minimum charge must accumulate in order to render the Q-ball energetically stable.
}

\graphicspath{{./figs/}}

\preprint{OUTP-21-30P}

\begin{document}

\tikzstyle{every picture}+=[remember picture]
\usetikzlibrary{shapes.geometric}
\usetikzlibrary{calc}
\usetikzlibrary{decorations.pathreplacing}
\usetikzlibrary{decorations.markings}
\usetikzlibrary{decorations.text}
\usetikzlibrary{patterns}
\usetikzlibrary{backgrounds}
\usetikzlibrary{positioning}
\tikzstyle arrowstyle=[scale=2]
\tikzstyle directed=[postaction={decorate,decoration={markings,
		mark=at position 0.6 with {\arrow[arrowstyle]{>}}}}]
\tikzstyle rarrow=[postaction={decorate,decoration={markings,
		mark=at position 0.999 with {\arrow[arrowstyle]{>}}}}]

\everymath{\displaystyle}

\maketitle

\section{Introduction}

Q-balls~\cite{Coleman:1985ki} are fascinating objects that can be found in theories of complex scalar fields. They are an example of a non-topological soliton (see Ref.~\cite{Lee:1991ax} and references therein) that are kept stable by a combination of conservation of energy and a Noether charge. Thus, Q-balls represent the state of minimum energy for a given charge. Though the theory is usually taken as that of a single complex scalar field symmetric under a group of $U(1)$ transformations, the principles have been extended to include multiple fields~\cite{Kusenko:1997zq}, more complicated symmetry groups~\cite{Safian:1987pr}, as well as gauging the stabilising symmetry~\cite{Lee:1988ag, Heeck:2021zvk}.

Finding the extremum of the energy functional is not analytically tractable in the most general case. However, two analytic limits exist, and represent the most studied examples of Q-balls. Specifically, these are the thin- and thick-wall limits~\cite{Coleman:1985ki, Kusenko:1997ad}. These are appropriate to describe Q-balls in the large and small charge limits, respectively. At least in the single-field case, not every theory that possesses a thin-wall limit has a corresponding thick-wall limit~\cite{PaccettiCorreia:2001wtt, Postma:2001ea, Lennon:2021uqu}. Recently, it has been shown that the thin-wall limit can be extended to describe Q-balls with smaller charges~\cite{Heeck:2020bau}.

Q-balls are not just interesting theoretical objects. Owing to their rich phenomenology, they have been studied in the context of real-world scenarios. In the absence of couplings to other light fields, Q-balls are absolutely stable, and so have been considered as candidates for dark matter~\cite{Kusenko:1997si, Kusenko:2001vu, Graham:2015apa, Ponton:2019hux}. Moreover, they have been shown to arise in supersymmetric theories~\cite{Kusenko:1997si, Kusenko:1997zq}, or in theories of extra dimensions~\cite{Demir:2000gj, Abel:2015tca}. At the time of writing, Q-balls have not been seen in any experiment, but it is believed that their signature will be striking if evidence for them is ever found~\cite{Gelmini:2002ez, Kusenko:1997vp, Croon:2019rqu, White:2021hwi}.

As mentioned above, the canonical Q-ball analysis has been extended to include multiple fields. However, the single-field scenario is still the most studied example. Given that extensions to the Standard Model often come with a plethora of scalar fields with non-trivial potentials, the multi-field scenario is relatively under-explored. In this paper, we consider the simplest multi-field example: a theory of a single complex scalar field and an arbitrary number of real scalar fields. The case of a non-topological soliton composed of a single complex scalar field with one additional real scalar field was first explored in Ref.~\cite{Friedberg:1976me}. It was noted that the real scalar, despite not carrying charge, would figure in the Q-ball if it allowed the resulting Q-ball to lower its energy either to one of greater stability, or even to achieve stability in the first place.

In this work, we review this original analysis according to the more modern approach as given in Ref.~\cite{Kusenko:1997zq}. Moreover, we also explore the thick-wall limit of the theory. As discussed in Refs.~\cite{PaccettiCorreia:2001wtt, Postma:2001ea, Lennon:2021uqu}, not every theory with a thin-wall limit possesses a thick-wall limit, which has ramifications for early universe phenomenology in solitosynthesis scenarios~\cite{Frieman:1989bx, Griest:1989bq, Kusenko:1997hj}. The added complexity introduced by considering multiple fields makes it impossible to derive an exact constraint in the same way. However, sufficient conditions can be found to restrict the theories to those with an analytic limit. Necessary conditions require a numerical study, which we leave to future work.

Specifically, in Section~\ref{sec:MultiFieldSingleSym:Minimisation}, we perform the minimisation procedure on the energy functional appropriate for the multi-field case, as first outlined in Ref.~\cite{Kusenko:1997zq}. In Section~\ref{sec:MultiFieldSingleSym:ThinWall}, we review the work of Ref.~\cite{Friedberg:1976me}, which considered thin-wall Q-balls comprising one charged scalar field with a single real scalar field, using the more modern approach of Ref.~\cite{Kusenko:1997zq}. In this section, we determine the differential equations that govern the vacuum expectation values (VEVs) of the complex and real scalars within the resulting thin-wall Q-balls, together with their physical properties of mass and volume. In Section~\ref{sec:MultiFieldSingleSym:ThickWall}, we analyse the thick-wall limit in the approximation scheme that the fields all have the same spatial profile up to a positive normalisation -- this is a necessity to make analytic progress. In reality, the spatial profiles of the fields might differ, but this extra freedom in the minimisation process can only further lower the Q-ball energy. Thus, any constraints found under this approximation scheme are sufficient to show the existence of thick-wall Q-balls, but are not strictly necessary. With this in mind, we show the existence of thick-wall Q-balls in a class of theories by demanding that the minimum be energetically stable against classical decay into the quanta of the charged scalar fields. We fully solve a special case -- one complex scalar with an arbitrary number of massless (or very light) scalars. Work with this assumption on the spatial profiles has been done in an ad-hoc manner in Refs.~\cite{Postma:2001ea, Bishara:2017otb}, but we seek to generalise these works here.

\section{Minimising the Energy in a Sector of Fixed Charge}
\label{sec:MultiFieldSingleSym:Minimisation}

We consider a theory of a single complex scalar, $\Phi (\vec{x},t)$, and $M$ real scalars, $\Psi_j(\vec{x},t)$. The Lagrangian density we consider is
\begin{equation}
\mathcal{L} = \partial_\mu \Phi \partial^\mu \Phi^* + \frac{1}{2} \sum^M_j \partial_\mu \Psi_j \partial^\mu \Psi_j - U(\Phi, \Phi^*, \Psi_j),
\end{equation}
where $U(\Phi, \Phi^*, \Psi_j)$ is some potential that we leave generic for now, stipulating only that it be a function of the fields, and not their derivatives, and that it vanish for vanishing field. The Euler-Lagrange equations for this theory are
\begin{equation}
\partial_{\mu}\partial^{\mu} \Phi + \frac{\partial U}{\partial \Phi^*} = 0 \quad \mathrm{and} \quad \partial_{\mu}\partial^{\mu} \Psi_j + \frac{\partial U}{\partial \Psi_j} = 0,
\end{equation}
with a similar equation governing the dynamics of $\Phi^*$.

We demand that the theory be invariant with respect to a single global $U(1)$ group of transformations defined through
\begin{equation}
\Phi \to e^{i \alpha} \Phi \quad \mathrm{and} \quad \Psi_j \to \Psi_j,
\label{eq:Symmetry}
\end{equation}
where $\alpha \in \mathbb{R}$ and the charge of the complex scalar is set to unity, with the real scalars uncharged. Associated to this symmetry is a Noether current density given by
\begin{equation}
\label{eq:MultiFieldSingleSym:MultiNoether}
j^{\mu} = i \left( \Phi \partial^{\mu} \Phi^{*} - \Phi^{*} \partial^{\mu} \Phi \right).
\end{equation}
This symmetry places a functional constraint on the potential, namely, that it be a function of the absolute value of the complex field.

A Q-ball is the state in a theory which minimises the energy for a fixed, non-zero Noether charge. The Hamiltonian for this theory is
\begin{equation}
H = \int \mathrm{d}^3x \left[ \dot{\Phi} \dot{\Phi}^* + \vec{\nabla} \Phi \cdot \vec{\nabla} \Phi^* + \frac{1}{2} \sum^M_j (\dot{\Psi}_j\dot{\Psi}_j+ \vec{\nabla} \Psi_j \cdot \vec{\nabla} \Psi_j) + U(\Phi, \Phi^*, \Psi_j)\right].
\end{equation}
To determine if this system admits Q-ball solutions, we introduce a Lagrange multiplier $\omega$, as in Ref.~\cite{Kusenko:1997zq}, that enforces charge conservation upon minimisation with respect to it:
\begin{equation}
\mathcal{E}_\omega = H + \omega \left( Q - \int\mathrm{d}^3 x \, j^0 \right),
\end{equation}
where $j^0$ is the zeroth component of the Noether current density given in Eq.~\eqref{eq:MultiFieldSingleSym:MultiNoether}. Thus, the functional we wish to analyse is given by
\begin{equation}
\begin{split}
\mathcal{E}_\omega = \omega Q + \int\mathrm{d}^3x & \left[ \dot{\Phi} \dot{\Phi}^* - i\omega \dot{\Phi}^* \Phi + i \omega \Phi^* \dot{\Phi} + \vec{\nabla} \Phi \cdot \vec{\nabla} \Phi^* \right.\\
& \left. + \frac{1}{2} \sum^M_j (\dot{\Psi}_j\dot{\Psi}_j+ \vec{\nabla} \Psi_j \cdot \vec{\nabla} \Psi_j) + U(\Phi, \Phi^*, \Psi_j) \right].
\end{split}
\end{equation}
We may complete the square on the first terms under the integral to give
\begin{equation}
\begin{split}
\mathcal{E}_\omega = \omega Q + \int\mathrm{d}^3x & \left[ \left|\dot{\Phi} - i \omega \Phi \right|^2 - \omega^2 \Phi^* \Phi + \vec{\nabla} \Phi \cdot \vec{\nabla} \Phi^* \right.\\
& \left. + \frac{1}{2} \sum^M_j (\dot{\Psi}_j\dot{\Psi}_j+ \vec{\nabla} \Psi_j \cdot \vec{\nabla} \Psi_j) + U(\Phi, \Phi^*, \Psi_j) \right].
\end{split}
\end{equation}
The terms containing derivatives with respect to time are the only terms with explicit time dependence. These are both positive semi-definite and are minimised if they each vanish. Thus, we require that
\begin{equation}
\label{eq:MultiFieldSingleSym:Ansatz}
\Phi (\vec{x}, t) = e^{i \omega t} \phi (\vec{x}) \quad \mathrm{and} \quad \Psi_j (\vec{x}, t) = \psi_j (\vec{x}),
\end{equation}
where $\phi (\vec{x})$ and $\psi_j (\vec{x})$ are functions purely of the spatial coordinate which we take, without loss of generality, to be real-valued. The energy functional is then
\begin{equation}
\label{eq:MultiFieldSingleSym:SpatialProfileFunc}
\mathcal{E}_\omega = \omega Q + \int\mathrm{d}^3x  \left[ \vec{\nabla} \phi \cdot \vec{\nabla} \phi - \omega^2 \phi^2 + \frac{1}{2} \sum^M_j ( \vec{\nabla} \psi_j \cdot \vec{\nabla} \psi_j) + U(\phi, \psi_j) \right].
\end{equation}
Notice that a spatial profile for the field $\phi$ that vanishes everywhere leads to a configuration of zero charge. Thus, a configuration of non-zero charge must have a spatial profile for the charged field that differs from zero in some finite domain -- this is not the case for the uncharged fields, which can be vanishing inside a multi-field Q-ball. The spatial profiles satisfy the equations
\begin{equation}
\nabla^2 \phi = \frac{1}{2} \frac{\partial}{\partial \phi} \left(U(\phi,\psi_j) - \omega^2 \phi^2\right), \quad \nabla^2 \psi_j = \frac{\partial U(\phi,\psi_j) }{\partial \psi_j}.
\end{equation}
These differential equations take the form of a bounce equation~\cite{Coleman:1977py, Callan:1977pt, Coleman:1977th}, with bounce potentials defined by $U(\phi,\psi_j) - \omega^2 \phi^2$ and $U(\phi,\psi_j)$, respectively. The lowest energy configurations are known to be spherically symmetric~\cite{{Coleman:1977th}}, under the boundary conditions of the field vanishing at spatial infinity, and being constant at the coordinate origin, and the first derivative vanishing at spatial infinity and the coordinate origin.

These highly-coupled differential equations cannot in general be solved analytically. However, under certain assumptions, analytic progress can be made. Namely, for stable Q-ball solutions to be found, we must have that $\omega_0 \leq \omega < m_\phi$, where $m_\phi$ is the mass of the complex scalar~\cite{Kusenko:1997ad}. The thin-wall limit corresponds to the limit that $\omega = \omega_0$ -- in fact, the thin-wall limit defines $\omega_0$. The thick-wall limit corresponds to $\omega \to m_\phi^-$.

\section{Thin-Wall Q-balls}
\label{sec:MultiFieldSingleSym:ThinWall}

The thin-wall limit for multi-field Q-balls was first considered in Ref.~\cite{Friedberg:1976me} for the case of one complex scalar and one real scalar.\footnote{It should be noted that in Ref.~\cite{Kusenko:1997zq}, the case of multiple charged scalars -- and no real scalars -- was analysed for thin-wall Q-balls.} For completeness, we include the analysis for multiple real scalars here. We note, however, that the physics is largely unchanged in going to this case.

In the thin-wall limit, the properties of the Q-ball are well-approximated by the properties of a spherical core of homogeneous Q-matter. We have that
\begin{equation}
\mathcal{E}_{\omega} \approx \omega Q + V\left[ U(\phi, \psi_j) - \omega^2 \phi^2 \right],
\end{equation}
where $V$ is the volume of the spherical core of the Q-ball. Minimisation with respect to the Lagrange multiplier yields
\begin{equation}
Q = 2\omega V \phi^2,
\end{equation}
as we would expect from the Noether current density given in Eq.~\eqref{eq:MultiFieldSingleSym:MultiNoether}. This expressions defines $\omega_0$, as stated above. Eliminating $\omega$ from $\mathcal{E}_{\omega}$ yields
\begin{equation}
E = \frac{Q^2}{4V \phi^2} + U(\phi, \psi_j) V.
\end{equation}
Minimising with respect to the volume gives us
\begin{equation}
V^2 = \frac{Q^2}{4 U(\phi, \psi_j) \phi^2},
\end{equation}
which, when eliminated, finally gives us an expression for the rest mass of a multi-field Q-ball:
\begin{equation}
m_Q = Q \sqrt{ \frac{U(\phi, \psi_j)}{\phi^2}}.
\end{equation}
This expression must be minimised such that the resulting Q-ball is classically stable against decay to the quanta of any of the fields $\phi$, i.e., that
\begin{equation}
m_Q < Q m_\phi
\end{equation}
where $m_\phi$ is the mass of the quanta of the field $\phi$. The minimisation procedure with respect to the field content leads to the conditions on the potentials that
\begin{equation}
\frac{\partial U(\phi, \psi_j)}{\partial \phi} = 2 \frac{U(\phi, \psi_j)}{\phi} \quad \mathrm{and} \quad \frac{\partial U(\phi, \psi_j)}{\partial \psi_k} = 0.
\end{equation}

A straightforward example of a potential that allows for multi-field Q-balls is the following:
\begin{equation}
U(\Phi^*,\Phi,\Psi) = m_\phi^2 \Phi^* \Phi - A\Psi\Phi^*\Phi + \lambda(\Phi^* \Phi)^2 + \lambda^{\prime}\Psi^4,
\end{equation}
where $\Phi$ is a massive, charged field (with charge set to unity) and $\Psi$ is a massless, uncharged field. The coefficients, $m_\phi^2$, $A$, $\lambda$, $\lambda^{\prime}$, are all taken positive, such that the sign assignment of each term is explicit. The Q-ball ansatz, given in Eq.~\eqref{eq:MultiFieldSingleSym:Ansatz}, means that we can rewrite this as 
\begin{equation}
U(\phi,\psi) = m_\phi^2 \phi^2 - A\psi\phi^2 + \lambda\phi^4 + \lambda^{\prime}\psi^4.
\end{equation}
The thin-wall conditions above yield the field values
\begin{equation}
\phi = \frac{1}{4} \frac{A}{(\lambda^3 \lambda^{\prime} )^{1/4}} \quad \mathrm{and} \quad \psi = \frac{1}{4} \frac{A}{(\lambda \lambda^{\prime} )^{1/2}}.
\end{equation}
The resulting Q-ball has a mass given by
\begin{equation}
m_Q = Q \sqrt{m_\phi^2 - \frac{1}{8}\frac{A^2}{(\lambda \lambda^{\prime} )^{1/2}}},
\end{equation}
which is clearly smaller than $Q$ amounts of the mass of the charged scalar, $m_\phi$. Thus, this theory allows for stable, multi-field Q-balls. Notice, there is special point where
\begin{equation}
\frac{1}{8}\frac{A^2}{(\lambda \lambda^{\prime} )^{1/2}} = m_\phi^2.
\end{equation}
In this case, the core of the thin-wall Q-ball does not contribute to the overall mass of the object -- it will all come from the wall. This was studied in the single-field case in Ref.~\cite{Spector:1987ag}

\section{Thick-Wall Q-balls}
\label{sec:MultiFieldSingleSym:ThickWall}

We now analyse multi-field Q-balls in the thick-wall limit. In order to analytically progress further, we must make an assumption about the spatial profile of the fields. This removal of a degree of freedom of the system leads to an upper bound on the mass of any stable Q-balls found, as additional freedom can only reduce the rest energy further. Thus, armed with this, we can derive sufficient conditions on the existence of thick-wall Q-balls on the class of theory of study. A full analysis for the necessary conditions would require a numerical analysis, which is outside the scope of this paper, and so we leave this to future work.

\subsection{The Spatial Profile Problem}

As noted in Ref.~\cite{Kusenko:1997ad}, the thick-wall limit is equivalent to the limit of small field values inside the Q-ball. Though this limit offers up the simplification that only the next-to-quadratic order in the potential is relevant, this is not enough to make analytic progress. The complication is purely due to the number of fields.

In Refs.~\cite{Postma:2001ea, Bishara:2017otb}, this problem was circumvented by assuming that all spatial profiles of the fields were the same up to a positive semi-definite normalisation constant for each field, with the resulting expressions minimised with respect to these normalisation constants.\footnote{In some sense, this is borrowing from the thin-wall analysis, whereby the field profiles are ignored apart from their homogeneous core values which differ only by a positive normalisation constant.} 
As stated above, the purpose of this ansatz is to provide sufficient conditions for the existence of thick-wall Q-balls in the class of multi-field theories of study, as well as to provide approximate values for the Q-ball properties. In reality, the spatial profiles of the fields might differ, but this extra freedom in the minimisation process can only further lower the Q-ball energy. To provide necessary conditions for a thick-wall limit, we must pursue a dedicated numerical analysis of the true spatial profiles of all individual fields, and this is the topic of future work. This would also allow us to see how accurate the analytic Q-ball properties are. Moreover, given that we have effectively reduced our multi-field system into one of a single-field, we expect to find results similar to those in Refs.~\cite{PaccettiCorreia:2001wtt, Postma:2001ea, Lennon:2021uqu}.

To be explicit, let the reference field be the charged scalar, $\phi$. As such, we define a set of positive semi-definite constants, $\{\beta_j\}$, such that
\begin{equation}
\psi_j = \beta_j \phi.
\end{equation}
The equations governing the spatial profiles of the scalars are now given by
\begin{equation}
\nabla^2 \phi = \frac{1}{2}\frac{\partial U}{\partial \phi} - \omega^2\phi \quad \mathrm{and} \quad \beta_j \nabla^2 \phi = \frac{\partial U}{\partial \psi_j}.
\end{equation}
Whether or not these equations are satisfied in general, and what it means for the properties of the Q-ball calculated, is the subject of future work. Nevertheless, we progress.




\subsection{The Thick-Wall Analysis}

The bounce potential in the thick-wall limit can now be written as
\begin{equation}
U_{\omega} (\phi, \beta_j) \approx \left[ \left( m_\phi^2 - \omega^2 \right)  +\frac{1}{2} \sum_j^M m_j^2 \beta_j^2 \right]\phi^2 - g(\beta_j) \phi^p,
\end{equation}
where $m_j$ is the mass of the $j$-th species of real scalar, and we take the next to quadratic term to be of order $p>2$. The coefficient of the quadratic term is clearly positive for $\omega < m_\phi$. The function $g(\beta_j)$ is model-dependent and so is left general for now, but it is assumed that it is positive definite, as otherwise there would be no Q-ball solution. The energy in Eq.~\eqref{eq:MultiFieldSingleSym:SpatialProfileFunc} is then
\begin{equation}
\begin{split}
\mathcal{E}_\omega = \omega Q + \int\mathrm{d}^3x &\left[ \left( 1 + \frac{1}{2} \sum^M_j \beta_j^2 \right) \vec{\nabla} \phi \cdot \vec{\nabla} \phi \right. \\
&\left. + \left[ \left( m_\phi^2 - \omega^2 \right)  +\frac{1}{2} \sum_j^M m_j^2 \beta_j^2 \right] \phi^2 - g(\beta_j) \phi^p \right].
\end{split}
\end{equation}
Consider the substitutions
\begin{equation}
\begin{split}
\xi_i & = \left(\dfrac{ \left( m_\phi^2 - \omega^2 \right) + \dfrac{1}{2} \sum_j^M m_j^2 \beta_j^2}{1 + \dfrac{1}{2} \sum^M_j \beta_j^2}\right)^{1/2} x_i,\\
&\\
\varphi & = \left(\frac{g(\beta_j)}{\left( m_\phi^2 - \omega^2 \right)  +\dfrac{1}{2} \sum_j^M m_j^2 \beta_j^2}\right)^{1/(p-2)}\,\phi.
\end{split}
\end{equation}
Then,
\begin{equation}
\label{eq:MultiFieldSingleSym:ThickWallEulerian}
\mathcal{E}_\omega = \omega Q +\dfrac{ \left( 1 + \dfrac{1}{2} \sum^M_j \beta_j^2 \right)^{3/2} \left[ \left( m_\phi^2 - \omega^2 \right)  +\dfrac{1}{2} \sum_j^M m_j^2 \beta_j^2 \right]^{(6-p)/(2p-4)}}{g(\beta_j)^{2/(p-2)}}S_\varphi, 
\end{equation}
where $S_\varphi$ is a dimensionless integral given by
\begin{equation}
S_\varphi = \int \mathrm{d}^3\xi \left[\vec{\nabla}_\xi \varphi \cdot \vec{\nabla}_\xi \varphi + \varphi^2 - \varphi^p\right].
\end{equation}
For different values of $p$, this has been numerically minimised in Ref.~\cite{Linde:1981zj}, with the general trend being that $S_{\varphi}$ increases for increasing $p$. Our expression for the energy must be minimised with respect to $\beta_j$ and $\omega$. To proceed further, we must specify the theory and therefore $g(\beta_j)$. However, we can restrict the types of theories that could plausibly lead to stable thick-wall Q-balls, and thus derive sufficient conditions for stable Q-balls to exist in the wider context of our assumption of the spatial profiles of the fields.

Consider the first derivative with respect to $\omega$. Requiring that this vanishes means that $p<6$ and leads to the condition
\begin{equation}
\epsilon = \Omega \left[ \left(1 - \Omega^2\right) + \frac{1}{2}\sum_j^M \beta_j^2 \mu_j^2\right]^{(10-3p)/(2p-4)},
\end{equation}
where
\begin{equation}
\epsilon \equiv \frac{Q}{S_{\varphi}} \left[\frac{p-2}{6-p}\right]  \dfrac{g(\beta_j)^{2/(p-2)}m_\phi^{(2p-8)/(p-2)}}{\left[ 1 + \dfrac{1}{2}\sum_j^M \beta_j^2 \right]^{3/2}},
\end{equation}
and we define the dimensionless parameters,
\begin{equation}
\mu_j \equiv \frac{m_j}{m_\phi} \quad \mathrm{and} \quad \Omega \equiv \frac{\omega}{m_\phi}.
\end{equation}
Note, $0<\Omega<1$, as $\omega<m_\phi$, by definition. From this condition, we can rewrite $\mathcal{E}_{\omega}$ as
\begin{equation}
\mathcal{E}_{\omega} = Qm_\phi \left[\Omega + \frac{1}{\Omega} \left(\frac{p-2}{6-p}\right) \left( \kappa - \Omega^2 \right)\right],
\end{equation}
where
\begin{equation}
\label{eq:MultiFieldSingleSym:kappa}
\kappa = 1 + \frac{1}{2}\sum_j^M \beta_j^2 \mu_j^2.
\end{equation}
Note, $\kappa \geq 1$ and is related to the mass of the configuration.\footnote{Technically, it is related to the mass to charge ratio of the configuration, but we have taken the charge of the complex scalar to be unity, and so this factor is not explicit.} Notice, $\kappa = 1$ if and only if $\beta_j = 0$ or $\mu_j = 0$ for each $j$. A subset of these requirements is the single-field case, and so it is entirely reasonable to expect that $\kappa = 1$ to have the same condition as for the single-field case. We show that this is indeed the case below. The regime in which $\kappa > 1$ is distinct from the case that $\kappa = 1$, as we see below.

For stable Q-balls to form, we require that $\mathcal{E}_{\omega} < Qm_\phi$. Thus,
\begin{equation}
\Omega + \frac{1}{\Omega} \left(\frac{p-2}{6-p}\right) \left( \kappa - \Omega^2 \right) < 1,
\end{equation}
which can be rewritten as
\begin{equation}
\label{eq:MultiFieldSingleSym:ConditionForStabilityp}
\frac{2(4-p)}{6-p}\Omega^2 - \Omega + \frac{p-2}{6-p} \kappa < 0.
\end{equation}
Notice, if $\Omega \to 0$, this condition is violated for $2<p<6$ and $\kappa>0$. However, since $\omega_0 \leq \omega < m_\phi$, this is acceptable. Notice, if $\kappa = 1$ this reduces to
\begin{equation}
(\Omega - 1)\left(\frac{2(4-p)}{6-p}\Omega - \frac{p-2}{6-p}\right) < 0,
\end{equation}
which is precisely the condition for single-field Q-balls as found in Ref.~\cite{Lennon:2021uqu}, i.e., that $2<p<10/3$. As stated above, the single-field case is an example of a theory for which $\kappa = 1$, and so this was expected. We must therefore now consider theories for which $\kappa > 1$.

For $\kappa > 1$ and $\Omega \to 1^{-}$, the condition given in Eq.~\eqref{eq:MultiFieldSingleSym:ConditionForStabilityp} must be satisfied. Considering $\Omega = 1$, this leads to
\begin{equation}
\frac{p-2}{6-p}(\kappa - 1)<0.
\end{equation}
We thus see that for any $\kappa > 1$, this condition is not satisfied. Thus, our requirement is broken at both $\Omega \to 0$ and $\Omega \to 1$. As such, if any stable Q-balls were to form, there would be a minimum charge required to reach stability. In order to determine if this can be the case, we require that both the roots of the quadratic given in Eq.~\eqref{eq:MultiFieldSingleSym:ConditionForStabilityp} lie in the range $0<\Omega<1$. The roots of this polynomial are
\begin{equation}
\Omega_{\pm} = \frac{6-p}{4(4-p)} \pm \frac{6-p}{4(4-p)} \sqrt{1 - 8\frac{(4-p)(p-2)}{(6-p)^2}\kappa}.
\end{equation}
For $\Omega\in(0,1)$, it is a necessary requirement that the first term on the right-hand side be in the range $(0,1)$, which leads to the condition
\begin{equation}
2<p<10/3.
\end{equation}
Requiring that the roots exist at all -- the term under the square root be positive -- leads to a $\kappa$-dependent condition for $p$,
\begin{equation}
1 - 8\frac{(4-p)(p-2)}{(6-p)^2}\kappa > 0.
\end{equation}
The function on the left-hand side is a monotonically decreasing function of $\kappa$ if $2 < p < 10/3$. Taking, therefore, the limit $\kappa \to \infty$ shows us that both roots always exist. Furthermore, the larger $\kappa$ is, the smaller the difference between the roots, and so the smaller the range of values which allow for a stable Q-ball.

To understand these requirements, we return to our definition of $\kappa$ as given in Eq.~\eqref{eq:MultiFieldSingleSym:kappa}. We note that we cannot form a stable Q-ball for all $Q>0$ unless the $j$-th species is either massless (or very light) or $\beta_j = 0$ for the species. A theory of this type ($\kappa = 1$) was the subject of Ref.~\cite{Bishara:2017otb}, whereby one real scalar was taken to be effectively massless, and the other had $\beta = 0$.

In the case that $\kappa>1$, we find that we have a lower bound on the charge of these Q-balls for them to be stable. This makes sense here -- the real scalars increase the mass of the Q-ball without increasing the charge of the Q-ball. Classical stability requires that the mass to charge ratio of the Q-balls be less than that of the constituent scalars. Thus, in the presence of real scalars, the Q-ball must accumulate enough charge to offset this increase in mass. This is therefore the origin of the requirements given above: as $\kappa$ gets larger, more charge must be accumulated, and the range of charges over which we have a stable, \textit{thick-wall} Q-ball decreases.


\subsection{A Worked Example: Arbitrary Number of Massless Real Scalars}

As stated above, the minimisation with respect to the $\beta_j$ is model-dependent and cannot be completed in general. As an example that can be fully worked out analytically, consider a simple theory of one complex scalar $\Phi$ with unit charge and $M$ massless scalars $\Psi_j$ defined through the Lagrangian
\begin{equation}
\begin{split}
\mathcal{L} = \,\,& \partial_\mu \Phi \partial^\mu \Phi^* + \frac{1}{2} \sum^M_j \partial_\mu \Psi_j \partial^\mu \Psi_j\\
& - \left( m_\phi^2 - \sum_j^M A_j \Psi_j \right) \Phi^*\Phi - \mathcal{O}\left(\Psi^4, (\Phi^*\Phi)^2, \Phi^*\Phi\Psi^2\right).
\end{split}
\end{equation}
Notice, we have taken the leading order interaction between the real and complex scalars to be cubic\footnote{Note, we could have also written down a term cubic in only the real scalars as it is not forbidden by the $U(1)$ symmetry. This class of term will only render the Q-ball more unstable as it will increase the energy (rest mass) of the configuration without increasing the charge. We omit this for simplicity here, but it is important to note.} -- this is because we are in the regime where $\kappa = 1$, and so $2<p<10/3$. We seek stable thick-wall Q-balls, and assume that all spatial profiles are, up to some positive semi-definite normalisation, the same. The functional in Eq.~\eqref{eq:MultiFieldSingleSym:ThickWallEulerian} is then given by
\begin{equation}
\mathcal{E}_\omega = \Omega m Q + m_\phi^3 \dfrac{ \left( 1 + \frac{1}{2} \sum^M_j \beta_j^2 \right)^{3/2} \left( 1 - \Omega^2 \right)^{3/2}}{\left( \sum_j^M A_j\beta_j \right)^2}S_\varphi,
\end{equation}
where $S_{\varphi} \approx 38.8$ after numerical minimisation~\cite{Linde:1981zj}. This expression must be minimised with respect to all $\beta_j$ and $\Omega$. First, we minimise with respect to all $\beta_j$. This process leads to the condition
\begin{equation}
\frac{3}{4}\beta_k = A_k \frac{\left( 1 + \frac{1}{2} \sum^M_j \beta_j^2 \right)}{\left( \sum_j^M A_j\beta_j \right)}.
\end{equation}
If the $k$-th species of real scalar does not have a coupling to the complex scalar that is cubic in the fields, it does not contribute to the Q-ball.\footnote{If a real scalar has a coupling to the complex scalar that is greater than cubic in the fields, it should contribute. However, this contribution to the properties of the Q-ball would be subleading to those calculated from only considering from the cubic coupling.} Taking $A_k$ to therefore be non-vanishing, we find
\begin{equation}
\frac{\beta_k}{A_k} = \frac{4}{3} \frac{\left( 1 + \frac{1}{2} \sum^M_j \beta_j^2 \right)}{\left( \sum_j^M A_j\beta_j \right)}.
\end{equation}
This expression holds for each species of real scalar, but the right-hand side is a constant for each species, and so we can we can then relate all species through
\begin{equation}
\frac{\beta_k}{A_k}  = \frac{\beta_l}{A_l}.
\end{equation}
We can use this to eliminate all but a single $\beta$ from the above condition since
\begin{equation}
\sum_j^M \beta_j^2 = \left(\frac{\beta_l}{A_l}\right)^2\sum_j^M A_j^2 \quad \mathrm{and} \quad \sum_j^M A_j \beta_j = \frac{\beta_l}{A_l} \sum_j^M A_j^2.
\end{equation}
We then find that
\begin{equation}
\beta_l = \dfrac{2A_l}{\left( \sum_j^M A_j^2 \right)^{1/2}}.
\end{equation}

Minimisation with respect to $\Omega$ yields
\begin{equation}
\epsilon = \Omega (1 - \Omega^2)^{1/2},
\end{equation}
where
\begin{equation}
\epsilon = \frac{4Q}{3^{5/2} S_{\varphi}m_\phi^2} \sum_j^M A_j^2,
\end{equation}
where we have eliminated all the $\beta_j$ by invoking the above condition. Notice, this has the same functional form as the single-field case given in Ref.~\cite{Kusenko:1997ad}. Thus, we may simply read off the solution that
\begin{equation}
\Omega^2 = \frac{1 + \sqrt{1 - 4\epsilon^2}}{2},
\end{equation}
where  $\epsilon < 1/2$. Given this, we may now expand $\mathcal{E}_{\omega}$ in powers of $\epsilon$ to give the mass of the Q-ball,
\begin{equation}
m_Q = m_{\phi} Q \left( 1 - \frac{1}{6}\epsilon^2 - \mathcal{O} \left( \epsilon^{4} \right) \right).
\end{equation}
We see that $m_Q < m_{\phi}Q$, and so the resulting Q-ball is indeed classically stable to decay to quanta of the field $\Phi$. The characteristic radius, $R$, and VEV of the field, $\langle \phi \rangle$, are given when $\xi \sim 1$ and $\varphi \sim 1$, and so
\begin{equation}
R^{-1} \sim \frac{\epsilon m_{\phi}}{\sqrt{3}} \left(1 + \mathcal{O}\left( \epsilon^2 \right) \right) \quad \mathrm{and} \quad \langle \phi \rangle \sim \frac{\epsilon^2 m_{\phi}^2}{2\left(\sum^M_j A_j^2\right)^{1/2}} \left(1 + \mathcal{O}\left( \epsilon^2 \right) \right).
\end{equation}
We thus see that these Q-balls are large compared to the Compton wavelength of the constituent complex scalars, and that the field inside the Q-ball is indeed small in some sense.

This analysis rested on two assumptions, namely, that the higher order term that stabilises the potential be small in comparison to the cubic and quadratic terms inside the Q-ball, and that $\mathcal{E}_{\omega}$ has a minimum in the range $0<\omega<m_{\phi}$. These conditions lead to constraints on the allowed charge of the thick-wall Q-ball solution. Assuming that this stabilising term comes in at order $q$ (as opposed to the minimal 4 given in the Lagrangian for this theory) and using the characteristic VEV given above, we demand that
\begin{equation}
\left(\sum_j^M A_j \beta_j \right) \langle \phi \rangle^3 \gg \frac{c_q}{\Lambda^{q-4}} \langle \phi \rangle^q,
\end{equation}
which leads to the constraint on $Q$,
\begin{equation}
Q \ll 604.8 m_{\phi} \left[ \frac{\Lambda^{q-4}}{c_q} \right]^{1/ (2q-6)}  \left[2 \left(\sum^M_j A_j^2\right)^{1/2}\right]^{(10 - 3q)/(2q-6)}.
\end{equation}
Demanding this for the quadratic term leads to the same constraint. The second condition on $Q$ arises from rearranging the condition $\epsilon<1/2$,
\begin{equation}
Q < 151.2 \frac{m_{\phi}^2}{\sum^M_j A_j^2}.
\end{equation}
Again, as in the single-field case, the upper bounds on the charge of these objects imply that these are small Q-balls.





\section{Summary}

The canonical Q-ball analysis is that of a single complex scalar field invariant under a $U(1)$ group of transformations. In this paper, we have reviewed the simplest multi-field generalisation, namely, that of a single complex scalar field together with an arbitrary number of real scalars. We found the physical parameters associated with the thin-wall Q-balls in this class of theory. This section was largely a review of the existing work found in Ref.~\cite{Friedberg:1976me}, albeit presented in a style similar to Ref.~\cite{Kusenko:1997zq}.

For the thick-wall limit, we presented a sufficient constraint for the existence of thick-wall Q-balls by assuming that all spatial profiles of the fields are the same up to positive semi-definite normalisation constant. To reiterate, these theories were classified by a parameter $\kappa\geq1$, defined in Eq.~\eqref{eq:MultiFieldSingleSym:kappa}. In summary, the case $\kappa = 1$, for which the single-field case is a subset, can possess energetically stable thick-wall solutions of any charge $Q>0$ when the index, $p$, of the next-to-quadratic term is in the range $2<p<10/3$. Similarly, for the case $\kappa > 1$, the index of the next-to-quadratic term is constrained to be in the range $2<p<10/3$, but is only energetically stable for a range of charges, where $Q_\mathrm{min} > 0$. This latter case arises as the real scalars add mass to the Q-balls, but do not reduce the charge, and so enough charge must accumulate in order for stability to be achieved.

The key part of the thick-wall analysis was the assumption that all fields have the same spatial shape. However, as we noted in the main text, this isn't necessarily valid from the point of view of the field equations. It is expected that the additional freedom afforded by the multitude of spatial profiles would allow for a lower energy solution to exist, and this is the fundamental reason for our work to represent sufficient, but not necessary, conditions. This assumption must be checked and verified numerically, with the necessary conditions for a stable thick-wall limit derived -- this will be the subject of future work.

\section*{Acknowledgments}

OL would like to thank John March-Russell for useful comments on this work, which was completed while OL was supported by the Colleges of St John and St Catherine, Oxford. Conversations with Fady Bishara on Q-ball physics over the years have also been particularly fruitful.

\bibliographystyle{JHEP}
\bibliography{paper}
\end{document}